# A Survey on Scheduling the Task in Fog Computing Environment

**FAIZA ISHAQ[1], HUMAIRA ASHRAF[2], NZ JHANJHI[3].**
[1,2] Department of computer and software Engineering international Islamic university
3 School of Computer Science, SCS, Taylors University
[1]Email: faiza.phdcs169@iiu.edu.pk
[2] Email: humaira.ashraf@iiu.edu.pk
3 Email: noorzaman.jhanjhi@taylors.edu.my

**ABSTRACT** With the rapid increase in the Internet of Things (IoT), the amount of data produced and bullet.png processed is also increased. Cloud Computing facilitates to handle storage, processing, and analysis of data as needed. However cloud computing devices are located far away from the IoT devices. Fog computing has emerged as a small cloud computing paradigm that is near to the edge devices and handles the task very efficiently. Fog nodes have a small storage capability than the cloud node but it is designed and deployed near to the edge device so that request must be accessed efficiently and executes in time. In this survey paper we have investigated and analysed the main challenges and issues raised in scheduling the task in fog computing environment. To the best of our knowledge there is no comprehensive survey paper on challenges in task scheduling of fog computing paradigm. In this survey paper research is conducted from 2018 to 2021 and most of the paper selection is done from 2020-2021. Moreover this survey paper organizes the task scheduling approaches and technically plans the identified challenges and issues. Based on the identified issues, we have highlighted the future work directions in the field of task scheduling in fog computing environment.

## I. INTRODUCTION

ROM the past few decades fog computing has introduced as a smart computing device that facilitates to handle and analyse the data as needed in a real time distributed internet of things (IoT) [1]. However, with the in-creasing amount of user request the amount of data exchange and travelling to the cloud is also increased in education and industrial fields [2][3]. whereas Cloud Computing devices are located far away from IoT devices and making use of network bandwidth, load balancing and implementation of virtualization [4]. Therefore Smart devices are increasing very speedily around the world that ultimately makes congestion in bandwidth consumption and latency. Cisco introduced the idea of fog computing and revealed that the IoT devices will enhance up to 30 billion by 2030 [5][6].The increased in latency and bandwidth consumption and delay in response time is not acceptable in critical situations like smart health systems and video surveillance etc. To overcome this issue, a fog node between cloud computing and IoT devices is introduced [7].The writer stated that [8] "Fog is nothing but Cloud that is nearer to the ground". Fog node manages the data closer to the edge devices and can optimally schedule the internet of things (IoT [8][9]. It provides facilities at the edge of a network efficiently [10]. The features of the fog computing helps the society to grow in different areas like active healthcare management, agriculture, waste and water management, and crucial traffic management system intelligently [11][12]. However, scheduling the resources in the fog computing paradigm is not a straightforward responsibility to be done easily. Scheduling means to efficiently assign the re-sources to the task and it is a hard optimization issue [13][7]. Task scheduling in fog computing is considered to be NP-hard. In the past two decades, several optimization algorithms are introduced to solve NP-hard problems [14] IoT task offloading to fog node significantly improve the QoS( Quality of Services), response time, load balancing, and energy consumption concurrently. Whereas in the case of cloud computing environment sending the request to the cloud and receiving back the result consumes a lot of bandwidth and delays the response time of the task as well. As a result, the middle layer called the fog computing layer is developed to overcome the limitations between cloud computing and IoT devices.fog computing having distributed nature facilitates to provide the cloud computing resources at the edge of the end devices [8][15][16][17][18][19][20][21][22][23] and [24]. However the memory capacity of the fog is less than the memory capacity of the Cloud therefore the fog performs task like cloud but in a limited capacity[25]. As the number of smart devices increases, the burden on fog node is also growing that may affect the load balancing in the fog computing environment. Therefore tasks should be scheduled in a fog computing environment in such a way that satisfy the QoS requirement of the tasks, balance the load among the fog node uniformly, and develop the improved response time and throughput of the task[26]. Hence by scheduling the task uniformly among all the nodes also improves and reduces the cost and energy consumption of the fog node and IoT devices significantly [27].

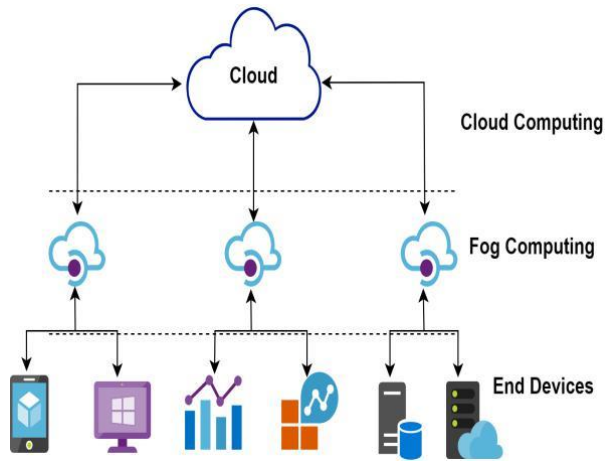

**FIGURE 1.** Fog Computing Architecture

Fog computing architecture is a three -layered architecture illustrated in fig.1. The top layer consists of cloud computing nodes. The intermediary layer called the fog computing layer is connected to the cloud layer and transfers the request to the cloud layer in case the requested task cannot be performed on the fog layer [28]. The bottom layer is the edge computing layer, which consists of IoT devices. The IoT devices send a request to the intermediate fog computing layer. Each fog node is a lightweight computing layer considered to be like a small cloud with the limited capability of storage, computation, and processing [29][30]. The main purpose of the fog node is to provide services near to edge node in a more efficient mode [31].

### A. MOTIVATION
The motivation of the paper is to uncover the challenges of scheduling in a fog computing environment in such a way that minimizes energy consumption while considering the load balancing of dependant and independent tasks simultaneously. Several optimized techniques have been applied to schedule the task and shows that real world problems are not considered in the development of scheduled environment. It also demonstrated that results may changes according to the scenarios. Most of the work is done on dependent task scheduling in homogeneous network instead of considering the independent task in heterogeneous network. From the comparison of the techniques, improvement can be made in modelling the technique that is more efficient and improves the performance of scheduling even in the heterogeneous network problems. These challenges and limitations are identified from the literature discussed below.

### B. CONTRIBUTIONS
1) However, as we know the challenges and issues arising in the cloud-fog computing environment. So, we have investigated and expressed various challenges in task scheduling approaches in fog computing environment. The purpose of this survey is to study the available issues in scheduling the task in fog computing environment. The main issues are expressed that are not analysed in a complete and systematic scenario. As a summary, the primary purpose of the paper is discussed below. Numerous scheduling approaches are proposed to overcome the problems raised in fog computing environment. Few of them are execution time, cost and transfer time of the task should be minimized, and load balancing among the fog nodes according to its capacity and energy consumption were also the main issues faced in fog computing environment.
2) Load balancing in centralized and distributed computing environment is very essential in scheduling the task in fog computing environment. Whereas it have been found that very few work is done on distributed computing environment.
3) We have also considered the load balancing along with the consumption of energy in a real world environment. Most of the techniques consider both load balancing and energy consumption simultaneously in the independent task scheduling environment.
4) Finally, we provide a literature review of task scheduling on fog computing environment, their research challenges in numerous areas, and identify various schemes applied to these research challenges.

The rest of the paper is organized as follows. Section 2 discusses the literature review. Comparative analysis is done in section 3, section 4 discusses Best Scheme from Each Literature Cluster and section 5 discusses the research gaps.

## II. LITERATURE REVIEW
Efficiently schedule the task in fog computing to improves the QoS requirements and minimizes the energy consumption can convert congested cities into smart cities. Different scheduling techniques have been discussed by the analyst and have tried to perform the real real-time scheduling problems efficiently. Some of the similar cluster-based techniques are discussed below.

### A. META HEURISTICS APPROACH
Ant Mating Optimization algorithm is proposed in paper [32] to schedule the task by minimizing the makespan and energy consumption in the fog computing environment. Since the demand for computational resources has been increased significantly from the last few decades in the fog environment, therefore, the proposed approach determines the best possible assignment of the given task based on the designed methodology function. In the paper, [32] scheduling is done by considering the ant model. Multiple jobs consist of a subset of tasks are deployed on the number of defined fog nodes in such a way that minimizes the total makespan and the energy consumption. The main emphasis of solving this issue is by adopting Ant Mating in nature. Ant Mating is a step-by-step process, in the initial step random number of ants is generated. Then the fitness value is calculated to select the number of queens having the highest fitness value or low-cost function values are considered and then the male ants that can crossover with the queens are selected. After crossover, mutation ultimately produces new offspring. After processing with the new offspring population, the new ant colony is created. The number of ant in a colony is predefined and if it exceeds the given number, the number of ants is destroyed until reach the maximum number of repetition (MaxRep). Scheduling is done following this ant mating approach. The proposed methodology is specifically designed for the energy consumption of the fog nodes, without considering the real-time task offloading of mobile devices. In future work, mobile users and network bandwidth optimization in dynamic and real-time task offloading can be considered to improve the performance in the real world scenario.

In the fog real-time computing environment the scheduling scheme changes after every interval of time. Scheduling the task to the appropriate fog node is an NP-hard problem. Therefore considering the dynamic event or the arrival of the new task, a dynamic workshop scheduling model is established based on Ant Colony System (ACS) [33]. The two dynamic scheduling methods namely parallel scheduling and parallel priority scheduling are designed correspondingly. Parallel scheduling is for the reduction of total makespan in the dynamic scheduling method whereas parallel priority scheduling reduces the receiving time of the events that are dynamic as well. Finally for the large problem set dual-objective dynamic job shop problem (DJSP) is established. In the paper the author proposed a parallel priority scheduling method will be based on urgency degree. The urgency degree is defined with the value of the urgency degree coefficient that determines the immediate emergency, general emergency and non-emergency. Selective-objective function for DJSP is based on this urgency degree coefficient value to choose the optimum scheduling method to meet the goal. The major purpose of this paper is to fulfil the requirement of the minimization of the makespan and deliver the task in time. If the task cannot deliver in time the scheduling method will cost the penalty.

ACS can be used to design the optimal solution of the problem to update the iteration to deal with selecting the best path at the interval of processing list of each node. The selective scheduling approach based on ACS and reduction in makespan based on DJSP is considered in the paper and concluded with the optimum results in case of dynamic event of different degree level of urgency. The selective scheduling strategy based on parallel and priority parallel scheduling achieve the best results depending on the situation to select the strategy. The efficiency of the algorithm degrades as the number of operations increases. It has been noticed that it is not appropriate for the extended time period in case of large scale scheduling problem. Therefore in future work the efficiency of the algorithm can be improved by considering the high operation efficiency.

Offloading the task efficiently in fog computing for IoT based application is proposed using Ant Colony Optimization (ACO) and Particle Swarm Optimization (PSO) approach [34]. The two meta-heuristic approaches are proposed to effectively schedule the load balance from the IoT devices to the fog nodes considering the communication cost and response time significantly. Moreover, the tasks of the IoT devices should be schedule to fog nodes in such a way that improves the QoS of IoT applications. The more the number of IoT devices and fog node upgrades the more the difficulties arises in handling the situation. Therefore to overcome the problem and achieve the goal ACO is considered to be the best scheduling approach. In the future work, two main points can be extended significantly. Firstly the optimization approach is required to consider the power consumption, communication and computation cost. Secondly the dynamic scenario also includes mobility in fog computing paradigm.

Fog computing intend to manage the task nearer to the end nodes so that optimizes the load over the network and improves the latency and response time. In order to efficiently organize and arrange the task to appropriate fog computing node, the researcher create Distributed Reconnaissance Ant Colony Optimization algorithm (DRACO) [35]. The algorithm is designed while considering the Multi-Objective Optimization problem and compared with the Single-Objective Optimization Problem taking into account both centralized and distributed synchronization. In the centralized system the communication and computation cost can make it an expensive system to build the ideal fog IoT computing net-work. Secondly failure occurs in any device results in the failure of the overall system. Therefore the distributed task placement is considered in the proposed ACO algorithm. Distributed network approach facilitate in increasing the efficiency, latency reduction and placement of the nodes to the appropriate device. The small scenario is build to compare the results of Multi-Objective optimization approach with the Single-Objective optimization algorithm. The DRACO algorithm is appropriate for the small scale scenario. For the large-scale scenario further revise criteria is required. Other limitations are the security system of discovering the shortest path secondly the load balancing should be implemented to schedule the task suitably.

A Privacy Aware Fog-enhanced Block chain-assisted task scheduling (PF-BTS) is introduced to improve the privacy system of the fog computing scheduling system [5]. An Ant Colony Optimization (ACO) algorithm in PF-BTS is designed to improve the latency measures and also the security and privacy of all the components of the system and give assurance of the location of the data; its identity and the important information are not exposed. The PF-BTS is compared with the Block chain based task scheduling protocol and shows high privacy awareness; this improves the total time of executing the task and system load significantly. The fog enabled environment ACO algorithm are deployed with the Block chain (BC) miners in order to optimize the assignment of the task in the VRs(Virtual Resource) of the cloud with the minimum execution time. Another problem taking into account is the privacy challenges, when it comes to schedule the task among cloud, fog and Iot devices. PF-BTS permits the fog node to perform the task execution at the end layer of the networks that increases the load balance but decreases the computational cycle performed by the VRs. While considering the privacy aware system model PF-BTS ignores the time taken to process the task. Therefore the future work is that, consumption of the time should be minimized in case of multiple processes executions.

Meta-heuristic approach called Whale Optimization algorithm (WOA) is introduced in the paper [36] to schedule task in such a way that minimizes the execution cost and reduce energy consumption significantly. WOA is inspired from the bubble net attacking technique called Humpback. Humpback whales create twisted or curved shape in the locality of the prey and enclose them. In a fog scheduling environment the fog node contains manager within the fog node to arrange the resources and the client requests within the Virtual Ma-chines (VMs) of the fog node. This improves the reduction of energy consumption considerably. WOA guarantees to search the most appropriate solution. Therefore it selects randomly the solution of the problems then in next iteration new search agent is reorganized in order to choose the best solution through randomly selecting the solution in all the iteration. This process continues until the satisfactory result is concluded. WOA algorithm is compared with the heuristics approach PSO, RR and SJF and results shows that WOA has the best performance in terms of energy consumption and cost. In future proposed WOA approach can be capable to work in real world problems with several objectives and optimizing the tasks.

Scheduling of the task is done with the help of energy aware marine predator algorithm (MPA) [10]. Further two more versions of MPA are proposed as: modified MPA (MMPA) and Improve MMPA. MPA consider the last updated position and improve MMPA consider the ranking strategy based on probability of re-initialization and predefined by the user. The half of the population is randomly re-initialize after a number of predefined iterations and perform the mutation in the last half towards the most excellent to this point. These three versions are proposed and calculate the performance metrics on the basis of consumption of energy, minimization of makespan and production rate of carbon dioxide and evaluated that Improve MMPA performs better than the other two versions. In the paper the task are considered as independent and are not associated with other tasks. Therefore future work of the paper is to apply MPA in case of dependant task in the fog computing environment as well.

A new task scheduling approach using meta-heuristic algorithm is proposed while considering reduction in energy consumption and minimizes the makespan in cloud- fog computing environment [37]. Therefore two techniques: one is the invasive weed optimization- cultural evolution algorithm (IWO-CEA) for energy aware task scheduling approach and second one is Dynamic Voltage and Frequency Scaling (DVFS) technique in order to reduce energy consumption in fog computing. Generally in such cases task scheduling has two main stages. First, arrange the task in a valid ordered sequence and then map the task to available computing resources. Second is DVFS technique, in order to make available suitable voltage and frequency supply for the fog computing devices. The proposed method is not applicable for the situation where deadline is not defined and also load balancing has been ignored while considering the resource utilization. Apart from that even as arranging the task in a sequence privacy and trust metrics were not measured. Therefore these metrics can be considered in a future work.

Cost-aware genetic-based (CAG) task scheduling algorithm is proposed for fog-cloud computing environment [38]. The proposed algorithm decides on the basis of cost reduction, whether the fog layer executes the request or the cloud layer executes it. The proposed algorithm also focuses on increasing the rate of responding the task successfully within the time deadline. Genetic based algorithm is proposed to run the system model and consist of a central scheduling node or a Master fog node whose responsibility is to schedule the task by estimating and calculating the time of finishing the task completely on the available fog-cloud nodes. If the resources are available it will be sent to the corresponding resources otherwise rejected. This process reduces the cost of searching the appropriate fog-cloud node and fulfils the task requests before the deadline. Genetic based algorithm (CAG) compares the financial cost and rate of success with

RR and Minimum Response Time algorithms and evaluated that CAG performs better. In the proposed algorithm task are considered to be independent and dependant task are ignored in the proposed scenario and secondly heavy workload of offloading the task is also need to be considered in future.

The on-line task scheduling for fog computing in NOMA (Non-orthogonal Multiple Access) based industrial IoT system is proposed for resource allocation and scheduling the task [39]. NOMA based fog computing (FC) system considers the task having multiple nodes and in case of immediate response sends the task to the closer helper node that improves the spectrum efficiency significantly. The main objective of the paper is to reduce the cost and energy consumption while taking into account the task scheduling and sub carrier allocation problem. Task scheduling and sub carrier allocation problem consists of binary variables; therefore on-line learning based optimization method is presented to tackle these challenging issues. In this paper an iterative algorithm is proposed in order to optimize the sub carrier allocation and task scheduling during the on-line learning process. Firstly, optimizing the power allocation and resource allocation in task scheduling is solved in the iterative algorithm then the sub carrier matching algorithm is proposed to minimize the cost effectively. The results show that proposed strategy performs better than using deep Q-learning approach. In this paper the proposed method has extended the model and considers the sub carrier allocation and matching theory algorithm in order to minimize the cost.

### B. MACHINE LEARNING

Several researches have been conducted to optimize the scheduling of the task in fog computing environment. The scheduling is done either on IoT devices or Fog nodes, scheduling at both end is rarely found in fog computing. Therefore to tackle such situation, this paper proposed a multi-criteria intelligent IoT Fog scheduling approach using game theory [40].Two main steps are applied in scheduling i.e. (1) based on minimizing the latency and resource utilization, preference list function for the IoT and preference list function for the fog layers are designed and ranked based on the set of criteria and (2) scheduling algorithm is designed for the centralized and distributed computing environment based on matching theory while considering the preferences of both IoT and fog node devices. This approach minimizes the latency of executing IoT services and also minimizes the makespan all together that ultimately improves the utilization of resources on the fog node. Three main metrics were considered while scheduling the task of IoT devices onto fog nodes: minimizing IoT task makespan, utilization of resources and execution time of the solution.

Fog computing overcome the limitation of cloud computing of managing the massive amount of data, specifically real time applications. This raises a new challenge of resource scheduling on the fog node and implements new strategy to guarantee the Quality of Services (QoS) requirement of applications. Therefore paper [41] emphasized on QoS requirement of application and proposed a strategy to categorize the applications into Classes of Services (CoS) on fog node. In the paper machine learning methodology is considered to categorize the classes to improve the QoS requirements. Fog infrastructure service makes the step by step process into packets. The first packet consists of applications as required by QoS and maps the application into CoS. The fog scheduler then uses the CoS to allocate resources according to their classes. The CoS is discriminated into seven different classes depending upon the user demands like military security, health care and hospital etc. are the critical and time constraints demands that need not to be delayed, similarly on the bases of user requests and scenario of the tasks CoS is categorized. Synthetic database is created relevant to the QoS requirement then the database is preprocessed and converts raw form of data into a usable data that can be used by Machine learning technique. ML algorithm puts the synthetic database into training and testing processes in order to calculate the efficiency of the CoS. This paper lacks to discuss some of the issues in fog network architecture which is included in the future work. For the future work more sensitive and critical demands are schedule to prioritize into the fog network scheduler.

Task allocation to best possible fog and cloud node is done in fuzzy clustering based task allocation approach in paper [42] to get the desired output based on minimizing the makespan, monetary and energy costs. Tasks in task allocation process are categorized as dependent and independent task. This dependent and independent task allocation is called multiple entity task allocation problems (META) To solve META problem for representing dependent and independent task, a new bipartite graph with fuzzy clustering task allocation approach called BiFTA is proposed. While considering the time delay, energy consumption and mandatory cost for selecting the most favourable node for each task. In addition the uncertainty executing problem is solved using fuzzy clustering and bipartite graph. In BiFTA approach, it is depended on a broker entity like having centralized computing nature, schedule and take decision, which may lead to overloaded or overcrowded and may affect the scheduling nature. Therefore an improved distributed nature paradigm should be designed by selecting multiple brokers based on defined criteria. And secondly each cloud-fog node contains different number of virtual machine, so that task allocation and processing is done in time.

A machine learning approach called Model Driven Software Engineering (MDSE) that enables and designs the IoT application to the most suitable fog node [27]. MDSE uses machine learning approach to places IoT applications at a most suitable fog node. Machine learning approach called rules based machine learning method is used to create rules based on user request after that it evaluate the capability of the fog node for the appropriate placement of the user request to the suitable fog node. This method minimizes the complexity of user request placement and provides an open free model that can be enhance and make changes according to the user's priorities and requirements changes. There are certain limitations in MDSE meta-model that it depends on the locations and requires a translator to translate in a required language and secondly it is not a user friendly as it requires implementation detail in the code. Therefore the future work can be that using the real time scenario this MDSE model can be further simulated to enhance the flexibility of the model. Secondly various customized scenarios of fog computing can also be considered during implementation and coding.

Fog computing nearest to IoE (Internet of Everything) helps to optimize the time delay sensitive task significantly. However to better schedule the task the author introduces the layer between IoE and fog nodes [43]. This layer integrate an intelligent and adaptive learning based scheduling technique having the capability of taking the decision of whether the incoming task should be process by fog or cloud. Adaptive and learning based approach called Learning Repository Fog Cloud (LRFC) is organized at the distributed gateways. LRFC layer schedule the task for fog or cloud depending upon the availability of the fog node. If all the fog nodes are occupied the scheduler send the request to the cloud node. After execution of the task the results are returned back to LRFC layer and LRFC Layer generates the result to IoE layer and updates its information accordingly. LRFC improves the QoS i.e response and processing time and energy consumption significantly. Load balancing among the fog and cloud node is not considered in LRFC layer and also this intelligent learning based technique should also be consider in large scale distributed computational network where fog, cloud and edge computing system are scheduled concurrently.

TABLE 1. critical analysis

| Paper | Year | Techniques | Proposed Techniques | Critical Analysis |
|---|---|---|---|---|
| (Hosseinioun et al. 2020b) | 2020 | Meta Heuristic Approach | Dynamic Voltage and Frequency Scaling (DVFS), hybrid Invasive Weed Optimization and Culture (IWO-CA) | Load balancing (Mekala et al. 2021), time deadline and privacy is ignored |
| (Ghanavati, J. H. Abawajy, and Izadi 2020b) | 2020 | | Ant Mating Optimization (AMO) | Mobile users and network bandwidth optimization in the dynamic and real time task offloading can be considered to improve the performance in the real world scenario. |
| (Hussein and Mousa 2020b) | 2020 | | Ant colony optimization (ACO) and particle swarm optimization (PSO) | Requires considering the power consumption and cost. secondly the dynamic scenario including mobility in fog computing paradigm should also be considered |
| (Abdel-Basset et al. 2020b) | 2020 | | Marine predators algorithm (MPA), meta-heuristic algorithms and genetic algorithms | Dependencies between the task is not considered and overburdening is still an issue (Mekala et al. 2021) |
| (K. N. Jayasena and Thisarasinghe 2019) | 2019 | | Whale optimization algorithm | real-world problems and independent task allocation is ignored |
| (Baniata, Anaqreh, and Kertesz 2021b) | 2021 | | Ant Colony Optimization (ACO) algorithm, Fog-enabled Blockchain-assisted scheduling model, namely PF-BTS. | Time delay is ignored |
| (Eyckerman et al. 2020c) | 2020 | | Distributed Reconnaissance Ant Colony Optimization algorithm | Security system of discovering the shortest path and load balancing should be considered. extremely slow for dynamic network (Eyckerman et al. 2020a). |
| (Y. Wan et al. 2020) | 2020 | | Event-driven dynamic workshop scheduling model - Ant Colony System (ACS) | does not apply for a long time in large scale scheduling problem |
| (K. Wang et al. 2020b) | 2020 | | Non-orthogonal multiple access (NOMA), On-line learning based algorithm | Arbitrary delay requirements are not considered that may affect the load balancing problem in fog nodes |
| (Nikoui et al. 2020b) | 2020 | | Cost-aware genetic-based (CAG) task scheduling algorithm | Dependant task are ignored in the proposed scenario and secondly heavy workload of offloading the task is also need to be considered |
| (Ghobaei-Arani et al. 2020b) | 2020 | | Moth-flame optimization algorithm, task scheduling approach | Load balancing is not considered |
| (Wu and Wang 2019b) | 2019 | Heuristic Based Algorithm | Deadline-aware Estimation of Distributed Algorithm (dEDA) with a repair procedure | Real time scenario and capacity of computing fog node is not considered |
| (Aburukba et al. 2020b) | 2020 | | Genetic algorithm (GA)- heuristic approach | maximize resource utilization and minimize latency |
| (Ghanavati, J. Abawajy, and Izadi 2020b) | 2020 | Machine Learning Approach | Dynamic Fault Tolerant Learning Automata (DFTLA) | Load balancing is not considered |
| (Guevara et al. 2020a) | 2020 | | Machine learning classification methodology | Real time system is not considered |
| (Arisdakessian et al. 2020b) | 2020 | | Multi-criteria intelligent IoT-Fog scheduling approach (game theory). | Energy consumption is not considered |
| (Ahmed. A. A. Gad-Elrab and Noaman 2019) | 2019 | | Bipartite graph, fuzzy clustering task allocation approach | Response time in a distributed computing nature is not measured |
| (Murtaza et al. 2020a) | 2020 | | Adaptive learning based task scheduling technique | Load balancing is not considered |
| (Shadroo, Rahmani, and Rezaee 2021b) | 2020 | | Deep learning methods, Self Organizing Map clustering Method(SOM) | Dependencies between the task and data transfer latency is not measured |
| (Gazori, Rahbari, and Nickray 2020b) | 2020 | | Double Deep Q-Learning (DDQL)-based scheduling algorithm | Average waiting time of the task is ignored |
| (Zaharia et al. 2020) | 2020 | | Offloading, greedy approach, multi-criteria optimization, genetic solution. | Assumes no Wi-Fi network is added and in turns no cost consumption is considered |
| (Arif et al. 2020b) | 2020 | | Model Driven Software Engineering (MDSE), fuzzy logics | Load balancing |

Paper [14] adopts deadline-aware estimation of distributed algorithm (dEDA) with repair procedure and local search engine to schedule the task and resource allocation order. Fog computing system schedules the IoT applications under the three layer system. First of all the probability based model is generated to plan and execute the problem. Then in the second step deDA algorithm with repair method and local search engine is propose to build. Last step conduct the experimental set-up to compare the efficiency of the whole algorithm, minimize the total tardiness of the task and meet the hard deadlines. The results are compared with the algorithm without repair procedure and local search and evaluated that the deDA is better than the simple algorithm without repair procedure and local search and perform much better in most of the cases. Load balancing in fog computing nodes and reality based system architecture should be considered in the future work to optimally schedule the task.

The research conducted in [28][44] considers energy consumption while scheduling the task in the fog computing environment. Several techniques are applied to reduce the energy consumption so that the overall cost should be minimized. These papers do not consider the privacy and trust metrics, tasks were considered to be independent, dependent tasks were not considered. Adding up, paper [37] addresses the energy aware method to reduce energy consumption. Priority based technique is applied to schedule the task, without considering the computational capacity and availability of the fog node, therefore load balancing problem occur while scheduling the resources do not consider the scheduling of the resources along with load balancing problem. Moth flame optimization approach is proposed in paper [45] to minimize the task execution and transfer time. Tasks are distributed equally among the fog nodes without considering the computing capacity of the fog node. Therefore the load balancing is also unobserved in this paper. The paper [46] proposed three cluster based methods i.e. Self Organizing Map (SOM), hierarchical SOM (H-SOM) and Auto-encoder SOM clustering method (AE-SOM). The results show that the proposed methodology improves the performance cost and bandwidth cost significantly. Dependent task were not consider in a proposed methodology. Double Deep Q-Learning (DDQL) based scheduling algorithm is proposed in Paper [47] in order to minimize the computation cost and service delay within the deadline constraints. This technique can further improve the scheduling mechanism by reschedule those task that have been waiting for the long time. This may reduce the average waiting time of the task accordingly. Makespan and delivery time is minimized by considering the two dynamic scheduling methods namely parallel scheduling and parallel priority scheduling [33].

Some of the optimization algorithm produces better results in energy consumption and makespan but at the same time ignoring to optimize the load balancing module. Pa-per [10][38][46] criticized on not considering the dependent tasks, they only considers the independent task. Load balancing in task scheduling is ignored in [27][37][43][45][48]. Paper [14][36][41] does not considers the real time scenario. Optimization done in paper [34][40] requires to optimize the power consumption problem. In our survey on scheduling tasks in Fog Computing environments, we draw upon foundational insights presented in [63-76]. Every technique has some advantages and limitation. Critical analysis of several techniques is illustrated in Table 1.

### III. COMPARATIVE ANALYSIS
Comparative analysis is done on the bases of objective based, performance based and technique based clustering.

#### A. OBJECTIVE BASED CLUSTERING
Objective based clustering defines the objective of every proposed technique. It is illustrated from the table 2 that most of the work is done on reducing energy consumption, makespan, and response time while scheduling the IoT devices in fog computing. Other parameters are resource management, load balancing, cost and QoS in a fog computing environment.

#### B. PERFORMANCE ANALYSIS
Schedule the task in fog computing environment in such a way that improves the QoS, response time, efficiency and total completion time of the task whereas at the same time cost factor is also measured in terms of reduction of energy consumption and the total cost of transferring and receiving the data. An intelligent task scheduling technique can better offload the task to the fog nodes while taking into account the load balancing among the fog nodes and minimizes the latency and average waiting time significantly. Table 3 explains the performance of several techniques. Papers [28][31][36][44] and [55] try to improve the energy consumption and ultimately the cost while scheduling the task in fog computing environment. The result shows that up to 92% less energy can be consumed. Similarly cost can be decrease up to 50 to 60% in a superlative scenario [31][36] and [52]. Other parameter describes in a tables 3 also reflect on the overall performance of the task scheduling system.

#### C. TECHNIQUE BASED ANALYSIS
Table 1 illustrates technique based clustering. Similar techniques are grouped together to form a cluster. Technique based clusters are based on meta-heuristic approach, machine learning approach, heuristic approach, artificial intelligence and mining. Most of the authors have used meta-heuristic approach and machine learning approach. Meta-heuristic approach is a problem independent technique and provides an optimal solution of the problem. Machine learning approach learns data from the information and then applies computational methods to compute the results.

**TABLE 2.** objective based clustering

| Paper | Year | Minimize total tardiness of task | Reduce energy consumption | Resource management | Reduce Execution/ makespan/ completion time | Load balancing | Response time | Maximize QoS | Privacy | Decrease cost | Efficiency | Task allocation | Minimizing the latency | Maximum data rate | Average waiting time/ latency |
|---|---|---|---|---|---|---|---|---|---|---|---|---|---|---|---|
| [14] | 2019 | ✓ | | | | | | | | | | | | | |
| [44] | 2020 | | ✓ | | | | | ✓ | | | | | | | |
| [27] | 2020 | | | ✓ | | | | | | | | ✓ | | | |
| [37] | 2020 | | ✓ | | | | | | | | | | | | |
| [49] | 2018 | | | | ✓ | | | | | | | | | | ✓ |
| [45] | 2020 | | | | ✓ | | | | | | | | | | |
| [32] | 2020 | | ✓ | | ✓ | | | | | | | | | | |
| [48] | 2020 | | ✓ | | | | ✓ | | | | | | | | |
| [34] | 2020 | | | | | ✓ | ✓ | | | | | | | | |
| [28] | 2020 | | ✓ | | | | | | | | | | | | ✓ |
| [10] | 2020 | | ✓ | | ✓ | | ✓ | | | | | | | | |
| [50] | 2018 | | ✓ | | | | | | | | | | | | |
| [31] | 2019 | | ✓ | | ✓ | | | | | | | | | | ✓ |
| [41] | 2020 | | | | | | | ✓ | | | | | | | |
| [36] | 2019 | | ✓ | | | | | | | ✓ | | | | | |
| [5] | 2021 | | | | | | | | ✓ | | | | | | |
| [51] | 2018 | | | | | ✓ | | | | ✓ | | | | | |
| [52] | 2020 | | | | | | | ✓ | | ✓ | | | | | |
| [53] | 2019 | | | | | ✓ | | | | | | | | | |
| [40] | 2020 | | | | ✓ | | | | | | ✓ | | | | |
| [42] | 2019 | | | | | | | | | | | ✓ | | | |
| [54] | 2020 | | | | | ✓ | | | | | | ✓ | | | |
| [9] | 2020 | | | | | | | | | | | | ✓ | | |
| [35] | 2020 | | | | | | | | | | | ✓ | | | |
| [43] | 2020 | | ✓ | | | | | ✓ | | | | | | | |
| [46] | 2020 | | | | | | | | | | ✓ | ✓ | | | |
| [47] | 2020 | | ✓ | | ✓ | | | | | | ✓ | | | | |
| [55] | 2020 | | ✓ | | | | | | | | ✓ | | | | |
| [56] | 2020 | | ✓ | | | ✓ | | | | | | | | ✓ | ✓ |
| [57] | 2020 | | | | | | | | | | ✓ | | | | |
| [58] | 2020 | | | | | ✓ | | | | | | | ✓ | | |
| [33] | 2020 | | | | ✓ | | ✓ | | | | | | | | |
| [59] | 2020 | | | | ✓ | ✓ | | | | | | | | | |
| [39] | 2020 | | | | | | | | | ✓ | | | | | |
| [60] | 2020 | | | ✓ | | | | | | | | | | | |
| [38] | 2020 | | | | | | | ✓ | | | ✓ | ✓ | | | |

## IV. BEST SCHEME FROM EACH LITERATURE CLUSTER

The research is conducted from 2018 to 2020 and researchers have identified several techniques. Similar techniques are grouped together to form a cluster. Three main clusters are assessing during the searching. First cluster is Meta-heuristic based cluster, second is heuristic based and third is machine learning. Each cluster is explained below in detail.

### A. META HEURISTIC APPROACH

Meta heuristic is an independent technique used in large scale of problems independently. During searching stage eleven papers were found related to the topic of interest. Following Meta heuristic approach several optimization techniques have been proposed. Ant Colony Optimization and Ant Mating Optimization is one of the frequently used technique other than whale optimization, invasive weed optimization, marine predator algorithm and particle swarm optimization techniques.

TABLE 3. Performance Analysis

| Paper | Year | Minimize total tardiness of task | Reduce energy consumption | Resource management | Reduce Execution/ makespan/ completion time | Load balancing | Response time | Maximize QoS | Privacy | Decrease cost | Efficiency | Task allocation | Minimizing the latency | Maximum data rate | Average waiting time/ latency |
|---|---|---|---|---|---|---|---|---|---|---|---|---|---|---|---|
| [44] | 2020 | | 16% | | | | 32% | | | | | | | | |
| [28] | 2020 | | 92.6% & 82.7% | | | | | | | | | | | | 85.29% & 67.4% |
| [31] | 2019 | | 11% | | 15.1% | | | | | 4.4% (network usage) | | | | | 7.78% |
| [36] | 2019 | | 4.47% & 4.50% | | | | | | | 62.07% & 60.91% | | | | | |
| [52] | 2020 | | | | | | | | | 58% | | | | | |
| [40] | 2020 | | | | 3to10 times | | | | | | 40% to 50% | | | | |
| [9] | 2020 | | | | 31% | | | | | | | | 21.9% to 46.6% | | |
| [55] | 2020 | | 85% | | | | | | | 7% | | | | | |

After a brief literature review it is evaluated that Ant Mating Optimization (AMO) technique [32] is the best scheme in Meta heuristic cluster. AMO resolve the issues of scheduling in fog computing environment and improves not only the completion time of the task but also improves the cost of energy consumption significantly. Best searching path or the best fog node for the task to process is easily calculated using AMO techniques.

### B. HEURISTIC APPROACH

Heuristic is a problem dependent approach and find a good enough solution to a specific problem. Heuristic approach is a small cluster of two papers based on genetic algorithm (GA) and deadline aware estimation of distributed algorithm. Among them genetic based algorithm improves the latency measures notably [9]. GA uses small scale of power and resources in fog computing paradigm and provides latency measure much better than using only Cloud computing services.

### C. MACHINE LEARNING

The second largest cluster is a machine learning based cluster. This cluster consist of nine papers and each methodology applied in a paper tries its best to evaluate the optimum result in terms of latency, load balancing, and energy consumption in fog computing environment. Mainly used techniques in machine learning are fuzzy clustering, game theory, adaptive learning based and deep learning based optimization approach. After a detailed literature review and result comparison evaluation, it is concluded that game theory [40] based evaluation gives better optimization approach then the others in terms of resource utilization and minimization of latency in fog computing environment.

### V. RESEARCH GAP

Literature review discussed above specifies many research gaps to be taken under consideration. In fog computing environment due to unpredictable nature of arrival of re-quested task, many research challenges are raised such as energy consumption, cost, load balancing, dependent and independent task placement, makespan, execution time, latency and average waiting time. The number of available resources and number of requests changes after every interval of time, therefore organizing such system that overcomes all challenges requires intelligent system architecture. Some of the research gaps are discussed below.

### A. COST

One of the most important research gaps to be taken under consideration is the cost of implementing a scheduled environment in a fog computing environment. Introducing the fog layer between the cloud and data users helps to reduce the communication cost significantly. On the other hand cost of executing the task in fog/cloud layer is a challenging issue to handle. That includes cost of bandwidth used while sending and receiving the task and the cost of using the computational resources [38].

**TABLE 4.** Technique based clustering

| Paper | Year | Techniques | Proposed Techniques |
|---|---|---|---|
| [37] | 2020 | | Dynamic Voltage and Frequency Scaling (DVFS), hybrid Invasive Weed Optimization and Culture (IWO-CA) |
| [32] | 2020 | | Ant Mating Optimization (AMO) |
| [34] | 2020 | | Ant colony optimization (ACO) and particle swarm optimization (PSO) |
| [10] | 2020 | | Marine predators algorithm (MPA), meta-heuristic algorithms and genetic algorithms |
| [36] | 2019 | | Whale optimization algorithm |
| [5] | 2021 | | Ant Colony Optimization (ACO) algorithm, Fog-enabled Blockchain-assisted scheduling model, namely PF-BTS. |
| [35] | 2020 | | Distributed Reconnaissance Ant Colony Optimization algorithm |
| [33] | 2020 | | Event-driven dynamic workshop scheduling model - Ant Colony System (ACS) |
| [39] | 2020 | | Non-orthogonal multiple access (NOMA), Online learning based algorithm |
| [38] | 2020 | | Cost-aware genetic-based (CAG) task scheduling algorithm |
| [45] | 2020 | Meta Heuristic Approach | Moth-flame optimization algorithm, task scheduling approach |
| [31] | 2019 | Multiobjective optimization approach | |
| [14] | 2019 | | Deadline-aware Estimation of Distributed Algorithm (dEDA) with a repair procedure |
| [9] | 2020 | Heuristic Based Algorithm | Genetic algorithm (GA)- heuristic approach |
| [56] | 2020 | | Dynamic Fault Tolerant Learning Automata (DFTLA) |
| [41] | 2020 | | Machine learning classification methodology |
| [40] | 2020 | | Multi-criteria intelligent IoT-Fog scheduling approach (game theory). |
| [42] | 2019 | | Bipartite graph, fuzzy clustering task allocation approach |
| [43] | 2020 | | Adaptive learning based task scheduling technique |
| [46] | 2020 | | Deep learning methods, Self Organizing Map clustering Method(SOM) |
| [47] | 2020 | | Double Deep Q-Learning (DDQL)-based scheduling algorithm |
| [56] | 2020 | | Offloading, greedy approach, multi-criteria optimization, genetic solution. |
| [27] | 2020 | Machine Learning Approach | Model Driven Software Engineering (MDSE), fuzzy logics |
| [49] | 2018 | Classification mining algorithm | I-Apriori, TSFC (Task Scheduling in Fog Computing) |
| [59] | 2020 | Swarm Intelligence(SI) | I-FASC, genetic algorithm(I-FA) |
| [53] | 2019 | Artificial Intelligence | Component-based scheduling algorithm |
| [28] | 2020 | Cache mechanism | Two energy-aware mechanisms, i.e., content filtration and load balancing |
| [44] | 2020 | | |
| [50] | 2018 | | Maximal energy-efficient task scheduling (MEETS) algorithm |
| [51] | 2018 | | Priority-based task scheduling algorithms |
| [52] | 2020 | | Real-time randomized algorithm (Power of Two Choices (Po2C) approach) |
| [54] | 2020 | | |
| [55] | 2020 | | Bag-of-Tasks workload model |
| [57] | 2020 | | Novel optimization technique, AMSGrad |
| [58] | 2020 | | Workload balancing scheme |
| [60] | 2020 | | Multi-agent based model |

### B. LOAD BALANCING

Fog layer is designed to process the tasks very immediately and responds very quickly but the storage capacity of the fog node is very less than the cloud node. Therefore task should be scheduled among the fog node in such a way that the load should be balanced among all the resources concurrently. Therefore load balancing in fog computing network is one of the challenge need to be cope intelligently [61].

### C. ENERGY CONSUMPTION

Energy consumption is another important challenge that comes under research gaps. With the increasing demands of resources on fog layer there is a need to design and increase the number of hardware devices. That ultimately increases the cost of energy consumption. Therefore tasks should be scheduled in a way that the minimum demand of hardware is requested that may reduce the energy consumption at fog layer significantly [61].

### D. SECURITY

Fog computing provides services to the edge devices more frequently and efficiently. However due to its distributed nature of paradigm it is in danger to security threats. Therefore security and privacy of the important data need to be process under the authenticated privacy policies [62].

### E. EXECUTION TIME

Schedule the task in fog computing environment is such a way that the execution time and response time of the tasks should be improved by assigning the task to the nearer fog computing node. The requests generated by the end user are executed in time more efficiently and accurately is the most important challenge of fog computing environment.

## References

[1] Redowan Mahmud et al. "Fog computing: A taxonomy, survey and future directions". In: Internet of Everything: Algorithms, Methodologies, Technologies and Perspectives (2018), pp. 103–130.

[2] Rajkumar Buyya et al. Cloud computing: Principles and paradigms. John Wiley & Sons, 2010.

[3] Marios D Dikaiakos et al. "Cloud computing: Distributed internet computing for IT and scientific re-search". In: IEEE Internet computing 13.5 (2009), pp. 10–13.

[4] Edra Resende de Carvalho et al. "The current context of Lean and Six Sigma Logistics applications in literature: A Systematic Review". In: Brazilian Journal of Operations & Production Management 14.4 (2017), pp. 586–602.

[5] Hamza Baniata et al. "PF-BTS: A Privacy-Aware Fog-enhanced Blockchain assisted task scheduling". In: Information Processing & Management 58.1 (2021), p. 102393.

[6] Dave Evans. "The internet of things: How the next evolution of the internet is changing everything". In: CISCO white paper 1.2011 (2011), pp. 1–11.

[7] Xin Yang and Nazanin Rahmani. "Task scheduling mechanisms in fog computing: review, trends, and perspectives". In: Kybernetes (2020).

[8] Flavio Bonomi et al. "Fog computing and its role in the internet of things". In: Proceedings of the first edition of the MCC workshop on Mobile cloud computing. 2012, pp. 13–16.

[9] Raafat O Aburukba et al. "Scheduling Internet of Things requests to minimize latency in hybrid Fog– Cloud computing". In: Future Generation Computer Systems 111 (2020), pp. 539–551.

[10] Mohamed Abdel-Basset et al. "Energy-aware marine predators algorithm for task scheduling in IoT based fog computing applications". In: IEEE Transactions on Industrial Informatics 17.7 (2020), pp. 5068–5076.

[11] Mandeep Kaur Saroa and Rajni Aron. "Fog computing and its role in development of smart ap-plications". In: 2018 IEEE Intl Conf on Parallel & Distributed Processing with Applications, Ubiquitous Computing & Communications, Big Data & Cloud Computing, Social Computing & Net-working, Sustainable Computing & Communications (ISPA/IUCC/BDCloud/SocialCom/SustainCom). IEEE. 2018, pp. 1120–1127.

[12] Heena Wadhwa and Rajni Aron. "Fog computing with the integration of internet of things: Architecture, applications and future directions". In: 2018 IEEE Intl Conf on Parallel & Distributed Processing with Applications, Ubiquitous Computing & Communications, Big Data & Cloud Computing, Social Computing & Networking, Sustainable Computing & Communications (ISPA/IUCC/BDCloud/SocialCom/SustainCom). IEEE. 2018, pp. 987–994.

[13] Brian P Gerkey and Maja J Mataric´. "A formal analysis and taxonomy of task allocation in multi-robot systems". In: The International journal of robotics research 23.9 (2004), pp. 939–954.

[14] Chuge Wu and Ling Wang. "A deadline-aware estimation of distribution algorithm for resource scheduling in fog computing systems". In: 2019 IEEE congress on evolutionary computation (CEC). IEEE. 2019, pp. 660–666.

[15] Subhadeep Sarkar et al. "Assessment of the suitability of fog computing in the context of internet of things". In: IEEE Transactions on Cloud Computing 6.1 (2015), pp. 46–59.

[16] Mazin Abdelbadea Nasralla Alikarar. "Scheduling IoT Requests to Minimize Latency in Fog Computing". PhD thesis. 2017.

[17] Maher Abdelshkour. "IoT, from cloud to fog computing". In: Cisco Blogs (2015).

[18] Soumya Kanti Datta et al. "Fog computing architecture to enable consumer centric internet of things services". In: 2015 International Symposium on Consumer Electronics (ISCE). IEEE. 2015, pp. 1–2.

[19] Mohammad Aazam and Eui-Nam Huh. "Fog computing and smart gateway based communication for cloud of things". In: 2014 International conference on future internet of things and cloud. IEEE. 2014, pp. 464–470.

[20] Shanhe Yi et al. "Security and privacy issues of fog computing: A survey". In: Wireless Algorithms, Systems, and Applications: 10th International Conference, WASA 2015, Qufu, China, August 10-12, 2015, Proceedings 10. Springer. 2015, pp. 685–695.

[21] Ivan Stojmenovic. "Fog computing: A cloud to the ground support for smart things and machine-to-machine networks". In: 2014 Australasian telecommunication networks and applications conference (ATNAC). IEEE. 2014, pp. 117–122.

[22] Shanhe Yi et al. "A survey of fog computing: concepts, applications and issues". In: Proceedings of the 2015 workshop on mobile big data. 2015, pp. 37–42.

[23] Clinton Dsouza et al. "Policy-driven security management for fog computing: Preliminary framework and a case study". In: Proceedings of the 2014 IEEE 15th international conference on information reuse and integration (IEEE IRI 2014). IEEE. 2014, pp. 16–23.

[24] Marcelo Yannuzzi et al. "Key ingredients in an IoT recipe: Fog Computing, Cloud computing, and more Fog Computing". In: 2014 IEEE 19th International Workshop on Computer Aided Modeling and Design of Communication Links and Networks (CAMAD). IEEE. 2014, pp. 325–329.

[25] Mandeep Kaur and Rajni Aron. "An energy-efficient load balancing approach for scientific workflows in

fog computing". In: Wireless Personal Communications 125.4 (2022), pp. 3549–3573.

[26] Mandeep Kaur and Rajni Aron. "Equal distribution based load balancing technique for fog-based cloud computing". In: International Conference on Artificial Intelligence: Advances and Applications 2019: Proceedings of ICAIAA 2019. Springer. 2020, pp. 189–198.

[27] Madeha Arif et al. "A model-driven framework for optimum application placement in fog computing using a machine learning based approach". In: Information and Software Technologies: 26th International Conference, ICIST 2020, Kaunas, Lithuania, October 15–17, 2020, Proceedings 26. Springer. 2020, pp. 102–112.

[28] Muzammil Hussain Shahid et al. "Energy and delay efficient fog computing using caching mechanism". In: Computer Communications 154 (2020), pp. 534–541.

[29] Ashkan Yousefpour et al. "Fog computing: Towards minimizing delay in the internet of things". In: 2017 IEEE international conference on edge computing (EDGE). IEEE. 2017, pp. 17–24.

[30] José Santos et al. "Fog computing: Enabling the management and orchestration of smart city applications in 5G networks". In: Entropy 20.1 (2017), p. 4.

[31] Mxolisi Mtshali et al. "Multi-objective optimization approach for task scheduling in fog computing". In: 2019 International Conference on Advances in Big Data, Computing and Data Communication Systems (icABCD). IEEE. 2019, pp. 1–6.

[32] Sara Ghanavati et al. "An energy aware task scheduling model using ant-mating optimization in fog computing environment". In: IEEE Transactions on Services Computing 15.4 (2020), pp. 2007–2017.

[33] Yi Wan et al. "Efficiency-oriented production scheduling scheme: an ant colony system method". In: IEEE Access 8 (2020), pp. 19286–19296.

[34] Mohamed K Hussein and Mohamed H Mousa. "Efficient task offloading for IoT based applications in fog computing using ant colony optimization". In: IEEE Access 8 (2020), pp. 37191–37201.

[35] Reinout Eyckerman et al. "Requirements for distributed task placement in the fog". In: Internet of Things 12 (2020), p. 100237.

[36] KPN Jayasena and BS Thisarasinghe. "Optimized task scheduling on fog computing environment using meta heuristic algorithms". In: 2019 IEEE International Conference on Smart Cloud (SmartCloud). IEEE. 2019, pp. 53–58.

[37] Pejman Hosseinioun et al. "A new energy-aware tasks scheduling approach in fog computing using hybrid meta-heuristic algorithm". In: Journal of Parallel and Distributed Computing 143 (2020), pp. 88–96.

[38] Tina Samizadeh Nikoui et al. "Cost-aware task scheduling in fog-cloud environment". In: 2020 CSI/CPSSI International Symposium on Real-Time and Embedded Systems and Technologies (RTEST). IEEE. 2020, pp. 1–8.

[39] Kunlun Wang et al. "Online task scheduling and resource allocation for intelligent NOMA based industrial Internet of Things". In: IEEE Journal on Selected Areas in Communications 38.5 (2020), pp. 803–815.

[40] Sarhad Arisdakessian et al. "FoGMatch: an intelligent multi-criteria IoT-Fog scheduling approach using game theory". In: IEEE/ACM Transactions on Net-working 28.4 (2020), pp. 1779–1789.

[41] Judy C Guevara et al. "On the classification of fog computing applications: A machine learning perspective". In: Journal of Network and Computer Applications 159 (2020), p. 102596.

[42] Ahmed AA Gad-Elrab and Amin Y Noaman. "Fuzzy clustering-based task allocation approach using bipartite graph in cloud-fog environment". In: Proceedings of the 16th EAI International Conference on Mobile and Ubiquitous Systems: Computing, Networking and Services. 2019, pp. 454–463.

[43] Faizan Murtaza et al. "QoS-aware service provision-ing in fog computing". In: Journal of Network and Computer Applications 165 (2020), p. 102674.

[44] Bushra Jamil et al. "A job scheduling algorithm for delay and performance optimization in fog computing". In: Concurrency and Computation: Practice and Experience 32.7 (2020), e5581.

[45] Mostafa Ghobaei-Arani et al. "An efficient task scheduling approach using moth-flame optimization algorithm for cyber physical system applications in fog computing". In: Transactions on Emerging Telecommunications Technologies 31.2 (2020), e3770.

[46] Shabnam Shadroo et al. "The two-phase scheduling based on deep learning in the Internet of Things". In: Computer Networks 185 (2021), p. 107684.

[47] Pegah Gazori et al. "Saving time and cost on the scheduling of fog-based IoT applications using deep reinforcement learning approach". In: Future Generation Computer Systems 110 (2020), pp. 1098–1115.

[48] Sara Ghanavati et al. "Automata-based dynamic fault tolerant task scheduling approach in fog computing". In: IEEE Transactions on Emerging Topics in Computing 10.1 (2020), pp. 488–499.

[49] Lindong Liu et al. "A task scheduling algorithm based on classification mining in fog computing environment". In: Wireless Communications and Mobile Computing 2018 (2018).

[50] Yang Yang et al. "MEETS: Maximal energy efficient task scheduling in homogeneous fog networks". In: IEEE Internet of Things Journal 5.5 (2018), pp. 4076–4087.

[51] Tejaswini Choudhari et al. "Prioritized task scheduling in fog computing". In: Proceedings of the ACMSE 2018 Conference. 2018, pp. 1–8.

[52] Farooq Hoseiny et al. "Using the power of two choices for real-time task scheduling in fog-cloud computing". In: 2020 4th International Conference on Smart City, Internet of Things and Applications (SCIOT). IEEE. 2020, pp. 18–23.

[53] Kenneth Johnson. Proceedings of the 6th IEEE/ACM International Conference on Big Data Computing, Applications and Technologies. Association for Computing Machinery, 2019.

[54] Georgios L Stavrinides and Helen D Karatza. "Scheduling a time-varying workload of multiple types of jobs on distributed resources". In: 2020 International Symposium on Performance Evaluation of Computer and Telecommunication Systems (SPECTS). IEEE. 2020, pp. 1–6.

[55] Dimitrios Tychalas and Helen Karatza. "A scheduling algorithm for a fog computing system with bag-of-tasks jobs: Simulation and performance evaluation". In: Simulation Modelling Practice and Theory 98 (2020), p. 101982.

[56] George-Eduard Zaharia et al. "Machine learning-based traffic offloading in fog networks". In: Simulation Modelling Practice and Theory 101 (2020), p. 102045.

[57] Xincheng Chen et al. "An energy-efficient mixed-task paradigm in resource allocation for fog computing". In: 2020 IEEE Wireless Communications and Net-working Conference (WCNC). IEEE. 2020, pp. 1–6.

[58] Qiang Fan and Nirwan Ansari. "Towards workload balancing in fog computing empowered IoT". In: IEEE Transactions on Network Science and Engineering 7.1 (2018), pp. 253–262.

[59] Shudong Wang et al. "Task scheduling algorithm based on improved firework algorithm in fog computing". In: IEEE Access 8 (2020), pp. 32385–32394.

[60] Fadoua Fellir et al. "A multi-Agent based model for task scheduling in cloud-fog computing platform". In: 2020 IEEE international conference on informatics, IoT, and enabling technologies (ICIoT). IEEE. 2020, pp. 377–382.

[61] Mandeep Kaur and Rajni Aron. "Energy-aware load balancing in fog cloud computing". In: Materials To-day: Proceedings (2020).

[62] R Elavarasi and Salaja Silas. "Survey on job scheduling in fog computing". In: 2019 3rd International Conference on Trends in Electronics and Informatics (ICOEI). IEEE. 2019, pp. 580–583.

[63] Shahid, H., Ashraf, H., Javed, H., Humayun, M., Jhanjhi, N. Z., & AlZain, M. A. (2021). Energy optimised security against wormhole attack in iot-based wireless sensor networks. Comput. Mater. Contin, 68(2), 1967-81.

[64] Wassan, S., Chen, X., Shen, T., Waqar, M., & Jhanjhi, N. Z. (2021). Amazon product sentiment analysis using machine learning techniques. Revista Argentina de Clínica Psicológica, 30(1), 695.

[65] Almusaylim, Z. A., Zaman, N., & Jung, L. T. (2018, August). Proposing a data privacy aware protocol for roadside accident video reporting service using 5G in Vehicular Cloud Networks Environment. In 2018 4th International conference on computer and information sciences (ICCOINS) (pp. 1-5). IEEE.

[66] Ghosh, G., Verma, S., Jhanjhi, N. Z., & Talib, M. N. (2020, December). Secure surveillance system using chaotic image encryption technique. In IOP conference series: materials science and engineering (Vol. 993, No. 1, p. 012062). IOP Publishing.

[67] Humayun, M., Alsaqer, M. S., & Jhanjhi, N. (2022). Energy optimization for smart cities using iot. Applied Artificial Intelligence, 36(1), 2037255.

[68] Hussain, S. J., Ahmed, U., Liaquat, H., Mir, S., Jhanjhi, N. Z., & Humayun, M. (2019, April). IMIAD: intelligent malware identification for android platform. In 2019 International Conference on Computer and Information Sciences (ICCIS) (pp. 1-6). IEEE.

[69] Diwaker, C., Tomar, P., Solanki, A., Nayyar, A., Jhanjhi, N. Z., Abdullah, A., & Supramaniam, M. (2019). A new model for predicting component-based software reliability using soft computing. IEEE Access, 7, 147191-147203.

[70] Gaur, L., Afaq, A., Solanki, A., Singh, G., Sharma, S., Jhanjhi, N. Z., ... & Le, D. N. (2021). Capitalizing on big data and revolutionary 5G technology: Extracting and visualizing ratings and reviews of global chain hotels. Computers and Electrical Engineering, 95, 107374.

[71] Nanglia, S., Ahmad, M., Khan, F. A., & Jhanjhi, N. Z. (2022). An enhanced Predictive heterogeneous ensemble model for breast cancer prediction. Biomedical Signal Processing and Control, 72, 103279.

[72] Zamir, U. B., Masood, H., Jamil, N., Bahadur, A., Munir, M., Tareen, P., ... & Ashraf, H. (2015, July). The relationship between sea surface temperature and chlorophyll-a concentration in Arabian Sea. In Biological Forum–An International Journal (Vol. 7, No. 2, pp. 825-834).

[73] Siddiqui, F. J., Ashraf, H., & Ullah, A. (2020). Dual server based security system for multimedia Services in Next Generation Networks. Multimedia Tools and Applications, 79, 7299-7318.

[74] Hanif, M., Ashraf, H., Jalil, Z., Jhanjhi, N. Z., Humayun, M., Saeed, S., & Almuhaideb, A. M. (2022). AI-based wormhole attack detection techniques in wireless sensor networks. *Electronics*, *11*(15), 2324.

[75] Jabeen, T., Jabeen, I., Ashraf, H., Jhanjhi, N., Humayun, M., Masud, M., & Aljahdali, S. (2022). A monte carlo based COVID-19 detection framework for smart healthcare. *Computers, Materials, & Continua*, *70*(2), 2365-2380.

[76] Hanif, M., Ashraf, H., Jalil, Z., Jhanjhi, N. Z., Humayun, M., Saeed, S., & Almuhaideb, A. M. (2022). AI-based wormhole attack detection techniques in wireless sensor networks. *Electronics*, *11*(15), 2324.